\documentclass{article}
\usepackage[a4paper,left=2cm,top=1.5cm,bottom=1.5cm,right=2cm]{geometry}
\usepackage[utf8]{inputenc}
\pdfoutput=1
\usepackage{color}
\usepackage{xcolor}
\usepackage{hyperref}
\definecolor{linkcolour}{rgb}{0,0.2,0.6}
\hypersetup{colorlinks,breaklinks,urlcolor=linkcolour, linkcolor=linkcolour, citecolor=blue}
\usepackage{graphicx}
\usepackage{caption}
\usepackage{subcaption}
\usepackage{multirow}
\usepackage{placeins}
\usepackage{booktabs}
\usepackage{authblk}
\usepackage{lineno}
\usepackage{afterpage}

\makeatletter         
\renewcommand\maketitle{
{
\begin{center}
{\huge \@title \par}
\vspace{1cm}
{\large \textbf{The ECFA Early-Career Researchers (ECR) Panel}}
\\[1cm]
{\large \@date}
\\[1cm]
\begin{minipage}{0.82\textwidth}
\normalsize The European Committee for Future Accelerators (ECFA) Early-Career Researcher (ECR) panel, which represents the interests of the ECR community to ECFA, presents in this document its initiatives and activities in the year 2023. This report summarises the process of the first big turnover in the panel composition at the start of 2023 and reports on the activities of the active working groups - either pursued from before or newly established. The overarching goal of the ECFA-ECR panel is to better understand and support the diverse interests of early-career researchers in the ECFA community and beyond.  
\end{minipage}
\end{center}
\vspace{1.5cm}
\begin{flushleft}
{The ECFA Early-Career Researchers (ECR) Panel: \href{mailto:ecfa-ecr-organisers@cern.ch}{ecfa-ecr-organisers@cern.ch}\\[0.5cm]\@author}
\end{flushleft}}}
\makeatother

\RequirePackage[style=numeric,backend=biber,sorting=none]{biblatex}
\begin{document}

\title{The ECFA Early-Career Researchers Panel: \\Report for the year 2023}
\date{\today}


\author[1]{Julia~Allen}
\author[2]{Bruno~Alves}
\author[*,3]{Jan-Hendrik~Arling}
\author[4]{Kamil~Augsten}
\author[5]{Emanuele~Bagnaschi}
\author[6]{Giovanni~Benato}
\author[7]{Anna~Bennecke}
\author[*,8]{Cecilia~Borca}
\author[9]{Paulo~Braz}
\author[10]{Lydia~Brenner}
\author[10]{Jordy~Degens}
\author[9]{Yannick~Dengler}
\author[11]{Christina~Dimitriadi}
\author[12]{Eleonora~Diociaiuti}
\author[5]{Laurent~Dufour}
\author[13]{Patrick~Dunne}
\author[14]{Ozgur~Etisken}
\author[5]{Silvia~Ferrario~Ravasio}
\author[15]{Nikolai~Fomin}
\author[10]{Andrea~Garcia~Alonso}
\author[16]{Leif~Gellersen}
\author[17]{Andreas~Gsponer}
\author[4]{Tomas~Herman}
\author[18]{Bojan~Hiti}
\author[19]{Laura~Huhta}
\author[*,20]{Armin~Ilg}
\author[21]{Kateřina~Jarkovská}
\author[22]{Jelena~Jovicevic}
\author[23]{Lucia~Keszeghova}
\author[24]{Henning~Kirschenmann}
\author[11]{Suzanne~Klaver}
\author[18]{Arman~Korajac}
\author[25]{Anastasia~Kotsokechagia}
\author[26]{Meike~Kussner}
\author[27]{Aleksandra~Lelek}
\author[28]{Guiseppe~Lospalluto}
\author[29]{Péter~Major}
\author[22]{Veljko~Maksimovic}
\author[30]{Jakub~Malczewski}
\author[31]{Carla~Marin~Benito}
\author[32]{Paula~Martinez~Suarez}
\author[33]{Vukasin~Milosevic}
\author[34]{Atanu~Modak}
\author[*,35]{Arnau~Morancho~Tarda}
\author[36]{Laura~Moreno~Valero}
\author[37]{Elisabeth~Niel}
\author[5]{Nikiforos~Nikiforou}
\author[18]{Anja~Novosel}
\author[19]{Petja~Paakkinen}
\author[*,38]{Holly~Pacey}
\author[39]{Rute~Pedro}
\author[*,20]{Marko~Pesut}
\author[25]{Guillaume~Pietrzyk}
\author[40]{Michael~Pitt}
\author[41]{Vlad-Mihai~Placinta}
\author[42]{Archita~Rani~Dash}
\author[17]{Géraldine~Räuber}
\author[43]{Mariana~Shopova}
\author[44]{Radoslav~Someonov}
\author[45]{Sinem~Simsek}
\author[46]{Kirill~Skovpen}
\author[47]{Filomena~Sopkova}
\author[39]{Fernando~Souza}
\author[*,48]{Elisabetta~Spadaro~Norella}
\author[*,49]{Marta~Urbaniak}
\author[50]{Lourdes~Urda~Gomez}
\author[16]{Erik~Wallin}
\author[51]{Valentina~Zaccolo}
\author[5]{Nima~Zardoshti}
\author[52]{Grzegorz~Zarnecki}

\affil[*]{Editor}
\affil[1]{University of Edinburgh, Edinburgh; United Kingdom}
\affil[2]{Ecole Polytechnique, Ile-de-France; France}
\affil[3]{Deutsches Elektronen-Synchrotron DESY, Hamburg; Germany}
\affil[4]{Faculty of Nuclear Sciences and Physical Engineering, Czech Technical University in Prague, Prague; Czech Republic}
\affil[5]{CERN, Geneva; Switzerland}
\affil[6]{INFN Laboratori Nazionali del Gran Sasso, L'Aquila; Italy}
\affil[7]{Université Catholique de Louvain, Brussels; Belgium}
\affil[8]{University of Torino and INFN, Turin; Italy}
\affil[9]{Institute for Physics, University of Graz, Graz; Austria}
\affil[10]{Nikhef National Institute for Subatomic Physics and University of Amsterdam, Amsterdam; Netherlands}
\affil[11]{Uppsala University, Uppsala; Sweden}
\affil[12]{National Laboratory of Frascati and INFN, Frascati; Italy}
\affil[13]{Blackett Laboratory, Imperial College London, London; United Kingdom}
\affil[14]{Kirrikale University, Kirrikale; Turkey}
\affil[15]{University of Bergen, Bergen; Norway}
\affil[16]{Lund University, Lund; Sweden}
\affil[17]{Institute of High Energy Physics, Austrian Academy of Sciences, Vienna; Austria}
\affil[18]{Department of Experimental Particle Physics, Jožef Stefan Institute and Department of Physics, University of Ljubljana, Ljubljana; Slovenia}
\affil[19]{Department of Physics, University of Jyväskylä and Helsinki Institute of Physics, University of Helsinki, Helsinki; Finland}
\affil[20]{Physik-Institut, University of Zürich, Zürich; Switzerland}
\affil[21]{Faculty of Mathematics and Physics, Charles University, Prague; Czech Republic}
\affil[22]{Institute of Physics, University of Belgrade, Belgrade; Serbia}
\affil[23]{Faculty of Mathematics, Physics and Informatics, Comenius University, Bratislava; Slovakia}
\affil[24]{Department of Physics, University of Helsinki, Helsinki; Finland}
\affil[25]{IRFU, CEA, Université Paris-Saclay, Gif-sur-Yvette; France}
\affil[26]{Institut für Experimentalphysik, Ruhr-Universität Bochum; Germany}
\affil[27]{Universiteit Antwerpen, Antwerpen; Belgium}
\affil[28]{Institute for Particle Physics and Astrophysics, ETH Zürich, Zürich; Switzerland}
\affil[29]{MTA-ELTE Lendület CMS Particle and Nuclear Physics Group, Eötvös Loránd University, Budapest; Hungary}
\affil[30]{Henryk Niewodniczanski Institute of Nuclear Physics Polish Academy of Sciences, Kraków; Poland}
\affil[31]{Institut de Ciencies del Cosmos, Universidad de Barcelona, Barcelona; Spain}
\affil[32]{Institut de Fisica d'Altes Energies, Universidad Autonoma Barcelona, Barcelona; Spain}
\affil[33]{Institute of High Energy Physics, Beijing; China}
\affil[34]{Rutherford Appleton Laboratory, Oxford; United Kingdom}
\affil[35]{Niels Bohr Institute, University of Copenhagen, Copenhagen; Denmark}
\affil[36]{Institut für Theoretische Physik, Westfälische Wilhelms-Universität Münster, Münster; Germany}
\affil[37]{École Polytechnique Fédérale de Lausanne, Lausanne; Switzerland}
\affil[38]{University of Oxford, Oxford; United Kingdom}
\affil[39]{Laboratório de Instrumentação e F\'isica Experimental de Part\'iculas - LIP, Lisboa; Portugal}
\affil[40]{Department of Physics, Ben-Gurion University, Beer-Sheva; Israel}
\affil[41]{Horia Hulubei National Institute of Physics and Nuclear Engineering, Bucharest-Magurele; Romania}
\affil[42]{Westfälische Wilhelms-Universität Münster, Münster; Germany}
\affil[43]{Institute for Nuclear Research and Nuclear Energy, Bulgarian Academy of Sciences, Sofia; Bulgaria}
\affil[44]{Sofia University, Sofia; Bulgaria}
\affil[45]{Istinye University, Istanbul; Turkey}
\affil[46]{Ghent University, Ghent; Belgium}
\affil[47]{Slovak Academy of Sciences, Bratislava; Slovakia}
\affil[48]{University of Genoa and INFN, Genoa; Italy}
\affil[49]{University of Silesia in Katowice, Katowice; Poland}
\affil[50]{Centro de Investigaciones Energ\'eticas Medioambientales y Tecnol\'ogicas (CIEMAT), Madrid; Spain}
\affil[51]{University of Trieste and INFN, Trieste; Italy}
\affil[52]{Henryk Niewodniczanski Institute of Nuclear Physics Polish Academy of Sciences, Kraków; Poland}

\maketitle

\clearpage
\section{Executive Summary}
The Early-Career Researchers (ECR) panel of the European Committee for Future Accelerators (ECFA)~\cite{ECFAECRPanel} formed in January 2021, following the recommendations of an initial ECR debate in November 2019~\cite{Bethani:2020ovr}, which aimed to provide ECR input to the 2020 update to the European Strategy for Particle Physics~\cite{EuropeanStrategyGroup:2020pow}.
Following this, the panel aims to continue to provide ECR input to ECFA, and to the 2026 update to the European Strategy for Particle Physics.

The ECR panel includes representatives from each ECFA member entity.
It is mandated to discuss all aspects that contribute in a broad sense to the future of the research field of particle physics, with an emphasis on topics of particular relevance to the ECR community.
The panel structure involves an organisation committee, responsible for organising and chairing meetings as well as handling outside correspondence, and working groups within the panel, which address or discuss particular topics in more detail.
A delegation of up to five members is selected from within the panel as observers to Plenary ECFA (PECFA) meetings, and one of these panellists is endorsed as an observer to Restricted ECFA (RECFA) meetings.

This annual report of the ECFA ECR panel aims to inform about the composition, the pursued activities and achievements of the ECR panel in 2023.
A similar report on the activities in 2021--2022 can be found in Ref.~\cite{report2021-2022}.
The structure of this report is as following: 
The membership composition and the active roles within the panel are introduced in Section~\ref{sec:panelstructure}. An overview of the active working groups and their activities are provided in Section~\ref{sec:wgs}, while Section~\ref{sec:communityint} summarises interactions between the ECR panel and the broader community in a variety of contexts. Section~\ref{sec:outlook} finally provides a short outlook for the panel into 2024 and beyond.

\section{Panel Structure}
\label{sec:panelstructure}
\subsection{ECR Panel Membership}



The ECR panel, as of December 2023, consists of 68 members representing 26 distinct entities: 25 ECFA member countries, plus CERN.
Each represented entity is allowed to have up to three members on the panel; countries hosting a major laboratory, as defined by being represented in the Lab Director's Group (LDG), are allowed a fourth representative, so long as at least one of the four representatives is from the laboratory in question.
As such, seven major laboratories have dedicated representation.

Panel members are selected by the national RECFA representatives, or the CERN RECFA representative, as appropriate.
It can also involve discussion amongst the appropriate PECFA representatives.
In some countries this follows an nomination/self-nomination procedure open to all ECRs in the country to obtain a list of candidates.
The selected members are then proposed to PECFA for endorsement: members endorsed by PECFA in November each year begin their term on January 1 of the following year, while members endorsed by PECFA in July each year begin their term on a back-dated date of July 1 of the same year.
The countries and laboratories represented in the ECFA ECR panel, and their representatives as of December 2023, are listed in Table~\ref{tab:currentpanel}.
An up-to-date list of members is maintained on the ECR panel section of the ECFA website~\cite{ECFAECRPanel}.

\begin{table}[h!]
\scriptsize
\begin{center}
\begin{tabular}{ llll }
  \toprule
  Country/Lab & ECR panel members & Position and speciality & Mandate Notes \\
  \midrule
\multirow{3}{*}{CERN}   & Emanuele Bagnaschi & Research Fellow, Theory & Ended 06/2023 \\
                        & Laurent Dufour & LD Research Fellow, LHCb & \\
                        & Silvia Ferrario Ravasio & TH Research Fellow, Theory & Started 07/2023 \\
                        & Nima Zardoshti & Staff, ALICE & \\
\midrule
\multirow{3}{*}{Austria}    & Yannick Dengler & PhD student, Theory & \\
                            & Andreas Gsponer & PhD student, Detector Physics & Started 07/2023 \\
                            & G\'{e}raldine R\"{a}uber & PhD student, HEPHY Vienna, Belle-II & \\
\midrule
\multirow{3}{*}{Belgium}    & Anna Bennecke & Postdoc, CMS & \\
                            & Aleksandra Lelek & Postdoc, Phenomenology & \\
                            & Kirill Skovpen & Postdoc, CMS & \\
\midrule
\multirow{2}{*}{Bulgaria}   & Mariana Shopova & Postdoc, CMS & \\
                            & Radoslav Simeonov & PhD student, ALICE & \\
\midrule
\multirow{1}{*}{Croatia}    & - & - & \\
\midrule
\multirow{1}{*}{Cyprus} & Nikiforos Nikiforou & Postdoc, ATLAS & \\
\midrule
\multirow{3}{*}{Czech Republic} & Kamil Augsten & Assistant Professor, ATLAS and COMPASS & \\
                                & Tomas Herman & PhD student, ALICE & \\
                                & Kate\v{r}ina Jarkovsk\'{a} & PhD student, theory & \\
\midrule
\multirow{1}{*}{Denmark}    & Arnau Morancho Tarda & PhD student, ATLAS & Started 07/2023 \\
\midrule
\multirow{3}{*}{Finland}    & Laura Huhta & PhD student, ALICE & \\
                            & Henning Kirschenmann & Senior Scientist, CMS & \\
                            & Petja Paakkinen & Postdoc, Theory & \\
\midrule
\multirow{2}{*}{France} & Anastasia Kotsokechagia & Postdoc, ATLAS & \\
                        & Guillaume Pietrzyk & Postdoc, LHCb & \\
\midrule
\multirow{3}{*}{Germany}    & Meike Kussner & Postdoc, PANDA/BESII & Started 07/2023 \\
                            & Laura Moreno Valero & PhD student, Theory & \\
                            & Archita Rani Dash & PhD student ATLAS & Started 07/2023 \\
\multirow{1}{*}{Germany/DESY}   & Jan-Hendrik Arling & Postdoc, ATLAS & \\
\midrule
\multirow{1}{*}{Greece} & - & - & \\
\midrule
\multirow{1}{*}{Hungary}    & P\'{e}ter Major & PhD student, CMS & \\
\midrule
\multirow{1}{*}{Israel} & Michael Pitt & postdoc, LHC/EIC Forward physics & \\
\midrule
\multirow{3}{*}{Italy}  & Cecilia Borca & PhD student, CMS &  \\
                        & Elisabetta Spadoro Norella & Postdoc, LHCb & \\
                        & Valentina Zaccolo & Assistant Professor, ALICE  &  \\
\multirow{1}{*}{Italy/INFN-LNGS}    & Giovanni Benato & Research Fellow, Experimental Neutrino Physics &  \\
\multirow{1}{*}{Italy/INFN-LNF} & Eleonora Diociaiuti & postdoc, Mu2e & \\
\midrule
\multirow{3}{*}{Netherlands}    & Jordy Degens & PhD student, ATLAS & \\
                                & Andrea Garcia Alonso & Postdoc, ATLAS & \\
                                & Suzanne Klaver & postdoc, LHCb &  \\
\multirow{1}{*}{Netherlands/NIKHEF} & Lydia Brenner & ATLAS staff, Nikhef & \\
\midrule
\multirow{1}{*}{Norway} & Nikolai Fomin & Posdoc, ATLAS & \\
\midrule
\multirow{3}{*}{Poland} & Jakub Malczewski & PhD student, LHCb & \\
                        & Marta Urbaniak & PhD student, NA61/SHINE & \\
                        & Grzegorz Zarnecki & Postdoc, T2K & Started 07/2023 \\
\midrule
\multirow{3}{*}{Portugal}   & Paulo Braz & Postdoc, Dark Matter & \\
                            & Rute Pedro & Postdoc, ATLAS/R\&D & \\
                            & Fernando Souza & PhD student, Phenomenology & \\
\midrule
\multirow{1}{*}{Romania}    & Vlad-Mihai Placinta & postdoc/electronics engineer, LHCb & \\
\midrule
\multirow{3}{*}{Serbia} & Jelena Jovicevic & Research Professor, ATLAS & \\
                        & Veljko Maksimovic & PhD student, ATLAS & \\
                        & Vukasin Milosevic & Postdoc, CMS & \\                
\midrule
\multirow{2}{*}{Slovakia}   & Lucia Keszeghova & PhD student, ATLAS & \\
                            & Filomena Sopkova & PhD student, ATLAS & \\
\midrule
\multirow{3}{*}{Slovenia}   & Bojan Hiti & postdoc, ATLAS & \\
                            & Arman Korajac & PhD student, Theory & \\
                            & Anja Novosel & PhD student, Belle II & \\
\midrule
\multirow{2}{*}{Spain}      & Carla Marin Benito & Assistant Professor, LHCb & \\
                            & Paula Martinez Suarez & PhD student, ATLAS & \\
\multirow{1}{*}{Spain/CIEMAT}      & Lourdes Urda Gomez & PhD student, CMS & \\
\midrule
\multirow{3}{*}{Sweden}     & Christina Dimitriadi & PhD student, ATLAS & Started 07/2023 \\
                            & Leif Gellersen & Postdoc, Phenomenology & Started 07/2023 \\
                            & Erik Wallin & PhD student, ATLAS/LDMX & Started 07/2023 \\
\midrule
\multirow{3}{*}{Switzerland}    & Armin Ilg & postdoc, FCC & \\
                                & Elisabeth Neil & Postdoc, LHCb & Started 07/2023 \\
                                & Marko Pesut & PhD student, Theory/Phenomenology & Started 07/2023 \\
\multirow{1}{*}{Switzerland/PSI}    & Giuseppe Lospalluto & PhD student, Muon physics & Started 07/2023 \\
\midrule
\multirow{2}{*}{Turkey} & Ozgur Etisken & Lecturer, FCC & Started 07/2023 \\
                        & Sinema Simsek & Postdoc, ATLAS & Started 07/2023 \\
\midrule
\multirow{3}{*}{UK} & Julia Allen & PhD student, ATLAS & \\
                    & Patrick Dunne & Lecturer, DUNE \& T2K & \\
                    & Holly Pacey & Research Fellow, ATLAS & \\
\multirow{1}{*}{UK/STFC-RAL}    & Atanu Modak & LHCb/CMS & \\
\bottomrule
\end{tabular}
\caption{The full list of ECFA ECR panel members over 2023.}
\label{tab:currentpanel}
\end{center}
\end{table}

\afterpage{\clearpage}

The official mandate of the ECR panel members was defined and approved by ECFA~\cite{mandate}.
This mandate defines the interactions between the panel and the parent ECFA group, including the allocation of up to five PECFA observers, one of which is also a RECFA observer; these will be discussed further in Section~\ref{subsec:PECFA}.
The mandate also notes that the panel meets infrequently; rather the overall activities are coordinated by an organisation committee discussed in Section~\ref{subsec:OC}, and day-by-day activities proceed in topical working groups, as detailed in Section~\ref{sec:wgs}.
Beyond defining a structure of the ECR panel, the mandate also defines the eligibility criteria for membership in the ECR panel and stresses that ``members act as individuals, but should be able to represent the views of early-career researchers in particle physics in the country from which they were nominated''.

The mandate serves as a solid foundation for the panel activities, but it is intentionally concise in order to leave the panel space to self-organise.
The panel members have therefore agreed upon a few further points in order to facilitate the activities of the ECR panel.
In particular, the panel has decided that quorum for any election, endorsement, or otherwise important choice requires at least 50\% of the panel members to have replied to the poll or other associated means of decision-making.
Additionally, the members have decided to hold at least three meetings of the panel each year: one in January, to set the priorities for the year; one around June, to prepare for the July PECFA meeting; and one around September, to prepare for the November PECFA meeting.
There is the option to have an additional fourth meeting; this fourth panel meeting was not deemed necessary in 2023.

While it is understood that not all panel members will be able to join any given meeting, the expectation is that all panel members will make a concerted effort to participate in the three/four panel meetings that take place each year.
The large time gap between panel meetings means that individual commitments often vary from one meeting to the next, which would support increased attendance over the course of the year, but this is not always the case: sometimes external commitments, such as coordination roles, necessitate attending recurring meetings on a long-term basis.
In previous years, the panel chose to constrain themselves to having each meeting on a different day of the week to the previous meeting in an attempt to avoid excluding panellists with consistent constraints from ever being able to join. 
However, in 2023, it was found to be better to just pick the meeting times to maximise attendance, as there was always one clearly favoured option among panellists.
Detailed minutes of the meetings are always circulated after the meeting, and a recording of the session as well as the presented slides are left on the Indico agendas, to allow any panellists who couldn't make the meetings to catch up.
Further to meetings or email communication, the panel opened a Mattermost server in 2023 to facilitate more efficient and informal discussion within the entire panel and within each working group and the organising committee.

\subsection{ECR Observers to Plenary and Restricted ECFA}
\label{subsec:PECFA}


According to the ECFA ECR panel mandate \cite{mandate}: ``From among the ECFA ECR Panel members, a delegation of up to five members is assigned by the panel as observers to Plenary ECFA meetings, and one member is assigned by the panel as observer to Restricted ECFA meetings''.
The following section presents the roles of Plenary and Restricted ECFA committees, responsibilities of observers, and the way how they were selected. 

PECFA's aim is to discuss and decide on all ECFA activities, including evaluating reports delivered by working groups, issuing recommendations to outside organisations, endorsing new country/laboratory representatives, and appointing the ECFA Chair and Secretary.
PECFA consists of approximately 100 members (for the full list see Ref. \cite{wwwPecfa}), who typically meet two times per year.
Each of these meetings consists of two parts: the first is closed and is used for discussions of topics relevant to ECFA functioning, while the second is open and focuses on informing the high energy physics community about recent activities carried out by ECFA.  

RECFA consists of a single representative from each member country/lab in ECFA (for the full list see Ref. \cite{wwwRecfa}).
This community serves as an advisory board to help the ECFA Chair and Secretary in shaping the programme of ECFA activities.
RECFA members also serve as the main point of contact with their respective local high energy physics communities and authorities.
RECFA meets a few times per year, including for country visits, after which the RECFA panellists compose a letter to the country's government and Particle Physics communities containing a set of recommendations.

The role and responsibilities of the ECR observers to PECFA and RECFA are as follows.
The observers take an active part in PECFA and RECFA meetings, representing the broad point of view of the ECR community.
They are also responsible for informing PECFA about activities carried out by the ECR panel.
In addition, they hold occasional meetings with the ECFA Chair and Secretary, with a purpose of discussing current developments and stimulating the work of the ECR panel.
The observers are also responsible for informing the ECR panel about ECFA activities, in particular those that require the direct involvement of the ECR panel or those that should be considered important by ECRs.
For example, the RECFA/PECFA observers will give a summary talk at each ECR panel meeting.

The list of the observers representing the ECFA ECR panel in the PECFA committee is shown in Table \ref{tab:rpecfaObservers}.

\begin{table}[!h]
\centering
\begin{tabular}{lll} 
\toprule
Name                 & Country/Lab & Role  \\ \midrule
Lydia Brenner        & Netherlands & PECFA/RECFA Observer \\
Eleonora Diociaiuti  & Italy       & PECFA Observer \\
Henning Kirschenmann & Finland     & PECFA Observer \\
Armin Ilg            & Switzerland & PECFA Observer \\
\bottomrule
\end{tabular}
\caption{The list of observers to the Plenary and Restricted ECFA groups in 2023.}
\label{tab:rpecfaObservers}
\end{table}

\subsection{ECR Organisation Committee}
\label{subsec:OC}

The role of the ECFA ECR Panel Organisation Committee (OC) is the overall coordination of the activities of the ECFA ECR panel and being the first point of contact.
Over the year 2023, three panel members were serving in the OC as listed in Table~\ref{tab:OCmembers}.

\begin{table}[!h]
\centering
\begin{tabular}{ll} 
\toprule
Name & Country/Lab \\ \midrule
Jan-Hendrik Arling & Germany/DESY \\
Holly Pacey        & UK \\ 
Valentina Zaccolo  & Italy \\
\bottomrule
\end{tabular}
\caption{The list of members of the ECFA ECR Panel Organisation Committee in 2023}
\label{tab:OCmembers}
\end{table}

In detail, the responsibilities for the OC are:
\begin{itemize}
    \item Organisation of the planned meetings of the ECR panel, including preparation of the agenda and its contributions, chairing the meetings, taking minutes;
    \item Point of contact for the ECFA secretary by handling requests from the ECFA chair and help maintain the public-facing website;
    \item Handling communication with the outside world by the publicly contactable ECFA ECR list (\href{mailto:ecfa-ecr-organisers@cern.ch}{ecfa-ecr-organisers@cern.ch});
    \item Arranging for public talk allocation as a ``speaker's committee'' type body;
    \item Coordination of the publication of panel reports like the ECR contribution to the bi-annual ECFA newsletters or the annual reports, such as this one.
\end{itemize}

\section{Panel Working Groups}
\label{sec:wgs}
For the effective working and the day-by-day activities, the ECR panel members are organising themselves within working groups for individual topics and events. These working groups are created on the basis of interests within the panel, where each member can voluntarily decide to take part in a group. Due to this setup it is naturally that the working groups, their members and activities are evolving with time, such that the content of covered topics is also dynamic.

In this section, the working groups with activities over the year 2023 are listed and a summary of their activities in 2023 are shown.

\subsection{Career Prospects and Diversity in Physics Programme Joint Survey}
\label{sec:wgs:cpdiv_survey}

In 2022, the Career Prospects and Diversity in Physics Programmes working groups jointly produced and distributed a comprehensive survey which aimed to better understand the issues faced by ECRs and what the ECFA ECR panel and wider HEP community to do to promote or solve these.
In March 2023, the survey was closed, and through the remainder of the year the results of the survey were analysed to produce a final report, which was published on the arXiv in April 2024 in Ref.~\cite{CPDivSurvey}.

\subsubsection{Working Group Composition}
The composition of the two working groups who have worked on the ECR survey analysis during 2023 is shown in Table~\ref{tab:survey_WG_composition}.

\begin{table}[htbp]
    \centering
    \begin{tabular}{ll}
    \toprule
    Name & Country \\\midrule
    Julia Allen & UK \\
    Giovanni Benato & Italy \\
    Kateřina~Jarkovská & Czech Republic \\
    Magdalena Kuich & Poland \\
    Aleksandra Lelek & Belgium \\
    Holly Pacey & UK \\
    Guillaume Pietrzyk & France \\
    G\'{e}raldine R\"{a}uber & Austria \\
    \bottomrule
    \end{tabular}
    \caption{Composition of the Career Prospects and Diversity in Physics working groups.}
    \label{tab:survey_WG_composition}
\end{table}

\subsubsection{Structure and circulation of the survey}

The survey was constructed to include around 100 questions assessing respondent demographics and their views and experiences in a wide variety of topics.
A summary of these is:
\begin{enumerate}
    \item Demographics.
    \item Experience of working in a research group.
    \item Experience of working in a collaboration.
    \item Views on the diversity of physics programmes.
    \item Experience and views on career prospects.
    \item Experience and views on Work-life balance.
    \item Experiences of discrimination.
    \item Views on respondents' recognition and visibility.
    \item Respondents' suggestions on what the ECFA ECFA Panel could do to support ECRs' career development.
\end{enumerate}

The survey was distributed among larger experiments, national mailing lists (via the ECFA national contacts and/or ECR panel members), and other ECR mailing lists.
The survey was also circulated further for the 2023 RECFA visit to the Czech Republic. 
Overall, 759 responses were obtained.

\subsubsection{Analysis results}

The survey analysis was divided into two parts: firstly clearly presenting the responses to each question in the survey, secondly to investigate how the responses were correlated between questions and how responses depended on respondent demographics.
Finally, the analysis results and respondents' answers to open questions asking for ideas of how the panel could help them, a set of recommendations was produced.
These suggest events which could be held to facilitate discussion in the academic particle physics community and further inform ECRs on a variety of topics.
Furthermore, it suggests actions that the wider HEP community could consider to improve ECR experiences.
Since the analysis results were extensive, rather than repeating them in this summary report we recommend that readers take a look at the full arXiv document~\cite{CPDivSurvey}.

\subsection{Software/Machine Learning applications for future colliders}
\label{sec:wgs:swml}

The working group is dedicated to software and machine learning for instrumentation and detector physics. The motivation for its creation was the analysis of a survey conducted by the instrumentation working group, which showed that the majority of people acknowledge the importance of software for their work in instrumentation; however, they report that they have not received any relevant training.

The previous and current activities of the group have focused on a comprehensive review of existing software/Machine Learning (ML) techniques and their applications in the field of instrumentation/detector physics and on preparing a survey aimed at better understanding the needs and challenges within the community. The overarching objective is to gather valuable insights that will not only serve as constructive feedback for existing schools on instrumentation but also play a central role in potentially designing a new school.

A pilot survey was released and is currently being circulated amongst our panel members and to the ECFA Training Panel for suggestions and input. The survey is structured starting with a general part with questions on the respondents' work experience, their personal interests and attendance of schools on Software and ML for instrumentation. The main body of the survey is then characterised by questions divided by subject and concerning the following topics of interest: Data Acquisition and Detector Control Systems, Detector Electronics, Detector Simulation and Machine Learning.  Once revised, the final survey will be circulated among the larger experiments, ECR and national mailing lists. The plan is to collect responses by the end of the summer and publish the results of the survey by the end of 2024. 

\begin{table}[htbp]
    \centering
    \begin{tabular}{ll}
    \toprule
    Name & Country \\\midrule
    Cecilia Borca & Italy \\
    Anastasia Kotsokechagia & France \\
    Elisabetta Spadaro Norella & Italy \\
    Marta Urbaniak & Poland \\     
      \bottomrule
    \end{tabular}
    \caption{Composition of the Software/ML applications for future colliders working group.}
    \label{tab:Software}
\end{table}

\subsection{Future Colliders}
\label{sec:wgs:fc}

The European Strategy for Particle Physics (ESPP) was last updated in 2020~\cite{EuropeanStrategyGroup:2020pow}. While strategy statement 3.a initiated the feasibility study of the Future Circular Collider (FCC), statement 7.b stressed the importance of ECRs in particle physics. Following this, ECFA created the ECFA ECR panel. Given this close connection of our panel to the ESPP and to the topic of future colliders, given the panel's name, the \textit{Future Colliders Working Group} (see Table~\ref{tab:future_colliders_WG_composition}) was created at the start of 2023.
This working group aims to inform ECRs on future colliders and foster discussions within the ECR community so that every ECR can shape their own opinion on our field's direction.

As our first objective, we organised the \textit{Future Colliders for Early-Career Researchers} event, which took place at CERN on the 24th September 2023 (\href{https://indico.cern.ch/event/1293507/}{https://indico.cern.ch/event/1293507/}). This hybrid ECFA-wide event attracted more than one hundred in-person participants, with a similar number joining the discussions over Zoom. The slides and recordings of the event are available on the Indico agenda and will serve as a cohesive reference for ECRs interested in future colliders. A summary of the event has furthermore just been published on the arXiv~\cite{FCreport}.

Many aspects of future colliders such as funding and career perspectives are country dependent. Using the knowledge gained during the organisation of the ECFA-wide event, we created a blueprint for national events on future colliders for ECRs \cite{blueprint}. The blueprint contains a list of sessions and talks that could make up such an event and links to previous talks given. The goal for 2024 is to have such national future collider events in as many ECFA countries as possible.

\begin{table}[htbp]
    \centering
    \begin{tabular}{ll}
    \toprule
    Name & Country/Lab \\ \hline
    Emanuele Bagnaschi & CERN \\
    Lydia Brenner & Netherlands \\
    Paulo Braz & Portugal \\
    Christina Dimitriadi & Sweden \\
    Patrick Dougan & UK \\
    Laurent Dufour & CERN \\
    Silvia Ferrario Ravasio & CERN \\
    Nikolai Fomin & Norway \\
    Armin Ilg & Switzerland \\
    Arman Korajac & Slovenia \\
    Giuseppe Lospalluto & Switzerland \\
    Arnau Morancho Tarda & Denmark \\
    Anja Novosel & Slovenia \\
    Petja Paakkinen & Finland \\
    Holly Pacey & UK \\
    Marko Pesut & Switzerland \\
    Michael Pitt & Israel \\
    Mariana Shopova & Bulgaria \\
    Erik Wallin & Sweden \\
    Nima Zardoshti & CERN \\
    \bottomrule
    \end{tabular}
    \caption{Composition of the Future Colliders working group.}
    \label{tab:future_colliders_WG_composition}
\end{table}

\subsection{Detector R\&D and Networking in Instrumentation WG}
\label{sec:detector_RnD}

This year no instrumentation-dedicated activities have been pursued in our panel. We plan to hold further editions of the Early-Career Instrumentation Forum (\href{ECIF}{https://indico.cern.ch/event/1203836/}), which was last held in October 2022. Future iterations of the ECIF could, for example, target specific instrumentation technologies or other topics relevant to ECRs such as training and networking in instrumentation.
On the topic of training, the ECFA Training Panel (\href{https://ecfa.web.cern.ch/ecfa-training-panel-collot-et-al}{https://ecfa.web.cern.ch/ecfa-training-panel-collot-et-al}) was formed, where also one representative from ECFA ECR is present.

\section{Summary of Community Interactions}
\label{sec:communityint}
Over the last year, the ECFA ECR panel has interacted with the wider high energy physics community in several different contexts.
Interacting with the wider ECR community in order to understand their priorities and views is fundamental to the panel's aims, in addition to conveying these to other groups. 
The following section summarises interactions this year.

As discussed in Section~\ref{sec:wgs:fc}, the panel organised the \textit{Future Colliders for Early-Career Researchers} event, which took place at CERN on the 24th September 2023, which had over one hundred in-person attendees.
The event summary published on the arXiv~\cite{FCreport}, and the slides/recordings on the agenda, will collectively form a great reference for the community.

As was discussed in Section~\ref{sec:wgs:cpdiv_survey}, the panel completed analysis of the Career Prospects and Diversity in Physics Programmes joint survey during 2023.
The analysis was written into a report published on the arXiv on the 2nd April 2024~\cite{CPDivSurvey}.
The report presents the views of 759 early-career researchers across the world, aiming to raise awareness of their experiences and challenges in academia to the wider high energy physics community.
The report also ended with sets of recommendations for what projects or events the panel could organise in future years, and for what actions the wider hep community could take to imprve the ECR experience.
Within 2024 we hope to present this report in various talks to ensure good engagement with the community.

At the EPS HEP 2023 conference in Hamburg, which took place from the 21st to the 25th August, Armin Ilg represented the panel by giving a talk on ``The ECFA Early-Career Researcher Panel''~\cite{EPSTalk}.
This talk introduced the panel, and summarised work done in 2021--2022, before introducing some of the work-in-progress topics from 2023.
Furthermore, at the 13th ICFA Seminar in DESY which took place from 27th November -- 1st December 2023, Lydia Brenner was invited to give a presentation on behalf of the panel titled ``Perspectives of Early-Career Scientists''~\cite{ICFATalk}.
This summarised the work done over 2023, with a focus on the outputs and future plans of the Future Colliders working group which is more relevant for ICFA.

As in previous years, the ECR panel submitted contribution to both ECFA Newsletters in 2023.
In the 11th Newsletter in the summer, this summarised work done within the first half of the year, presenting some preliminary results from the Career Prospects and Diversity in Physics Programme Joint Survey and the talk given at the EPS-HEP 2023 conference.
In the 12th Newsletter, we summarised the release of the first part of the survey analysis document, the successful ``Future Colliders for Early-Career Researchers'' event, and the presentation from our panel given at the ICFA Seminar.
All of the newsletters can be found in Ref.~\cite{newsletters}.

Anyone from the high energy physics community is always welcome to contact the ECR panel's organisation committee (\href{mailto:ecfa-ecr-organisers@cern.ch}{ecfa-ecr-organisers@cern.ch}) to suggest future activities or other types of interactions that should be pursued.
Anyone who is interested in following the activities of the ECR panel is welcome to sign up to the announcement mailing list (\href{mailto:ecfa-ecr-announcements@cern.ch}{ecfa-ecr-announcements@cern.ch}); this is a CERN-hosted mailing list, and thus registration requires either having a CERN account or CERN lightweight account, the latter of which does not require any CERN affiliation.

\section{Outlook}
\label{sec:outlook}
The year 2023 was, for the ECFA ECR panel, a year of sizeable transition, due to a large turnover of panellists whose 2-years mandates ended.
Several members stayed on the panel with renewed mandates to aid the transition phase, and led a critical review on the self-given organisation and structure.
It was decided to keep the existing structure, but to evolve the work focus onto relevant topics which are timely or proposed by the panellists.
As such, some working groups were not pursued, or were less actively worked on, compared to the panel's first two years of activity.
Instead, new working groups were created and already delivered remarkable output for the broader community concerning the interests of ECRs in the field.

Preparation for the next update of the European Strategy will start in 2024.
In line with the panel's mandate, it will promote further discussion of topics like future colliders amongst ECRs (for example by organising events following the national event blueprint), and actively participate in existing planned discussions.
We will also aim to publish the results of the survey on Training Quality in Instrumentation Software and Machine Learning.
Moreover, there will be another large turnover of panellists, leading to a panel almost orthogonal in membership to the original composition.

The activities of the ECR panel in 2023 have continued to enable the views, concerns, and ideas of ECRs to influence discussions and decision making across a broad range of areas of high energy physics.
This report has highlighted the main achievements of the panel in 2023, and has provided a brief look to the future challenges and opportunities for the incoming ECR panellists.

\clearpage
\addcontentsline{toc}{section}{Bibliography}
\printbibliography[title=References]

\end{document}